\documentclass[%
 reprint,
 superscriptaddress,
 amsmath,amssymb,
 aps,
 floatfix,
 showkeys
]{revtex4-1}
\usepackage{graphicx}
\usepackage{dcolumn}
\usepackage{bm}

\usepackage{xcolor}

\usepackage{setspace}

\begin{document}

\preprint{APS/123-QED}

\title{Spatially resolved dielectric loss at the Si/SiO\textsubscript{2} interface}

\author{Megan Cowie}
\affiliation{Department of Physics, McGill University, Montr\'eal, Qu\'ebec, Canada}
\email{megan.cowie@mail.mcgill.ca}
\author{Taylor J.Z. Stock}
\affiliation{London Centre for Nanotechnology, University College London, London, United Kingdom}
\affiliation{Department of Electronic and Electrical Engineering, University College London, London, United Kingdom}
\author{Procopios C. Constantinou}
\affiliation{London Centre for Nanotechnology, University College London, London, United Kingdom}
\author{Neil J. Curson}
\affiliation{London Centre for Nanotechnology, University College London, London, United Kingdom}
\affiliation{Department of Electronic and Electrical Engineering, University College London, London, United Kingdom}
\author{Peter Gr\"{u}tter}
\affiliation{Department of Physics, McGill University, Montr\'eal, Qu\'ebec, Canada}

\date{\today}

\begin{abstract}
The Si/SiO\textsubscript{2} interface is populated by isolated trap states which modify its electronic properties. These traps are of critical interest for the development of semiconductor-based quantum sensors and computers, as well as nanoelectronic devices. Here, we study the electric susceptibility of the Si/SiO\textsubscript{2} interface with $\mathrm{nm}$ spatial resolution using frequency-modulated atomic force microscopy. The sample measured here is a patterned dopant delta-layer buried $2~\mathrm{nm}$ beneath the silicon native oxide interface. We show that charge organization timescales of the Si/SiO\textsubscript{2} interface range from $1-150~\mathrm{ns}$, and increase significantly around interfacial traps. We conclude that under time-varying gate biases, dielectric loss in metal-insulator-semiconductor (MIS) capacitor devices is in the frequency range of $\mathrm{MHz}$ to sub-$\mathrm{MHz}$, and is highly spatially heterogeneous over $\mathrm{nm}$ length scales.
\end{abstract}

\maketitle

Semiconductors are emerging as a promising platform for spin-based quantum sensing and computation, with a clear path to scalability and long coherence times. In one widely adopted architecture, single dopant atoms are buried some nanometers beneath the semiconductor surface, where they are electronically accessed by means of an applied gate voltage\cite{Wellard2003, Morton2011, Zhang2018, GonzalezZalba2021}. Silicon is a promising host lattice, in large part because existing Si microfabrication technologies are unparalleled for any other material\cite{Morton2011,GonzalezZalba2021}. However, it is impossible to fabricate a Si surface which is entirely homogeneous: In particular, if the surface has a SiO\textsubscript{2} overlayer, a variety of defects such as interfacial traps (ITs, such as $P\textsubscript{b0}$ and $P\textsubscript{b1}$ centres\cite{Kato2006}) populate the Si/SiO\textsubscript{2} interface, modifying the surface electronic environment and resulting in, for example, random telegraph fluctuations (1/f noise)\cite{Kirton1989,Payne2015,Kirton1989b,Fleetwood2023} and threshold voltage shifts\cite{Campbell2007}. 

ITs in silicon devices have been studied using a wide range of techniques\cite{Fleetwood2008,Engstrom2014}, including transport measurements\cite{Koh1997,Fleetwood2023}, capacitance\cite{Engstrom1983} and admittance spectroscopy\cite{Nicollian1967, Raeissi2010,Raeissi2008,Li2022,Piscator2009,Li2021,Yldz2011,BirkanSeluk2007}, deep level transient spectroscopy\cite{Engstrom1983}, photoemission spectroscopy\cite{Moritz2023}, and electron spin resonance spectroscopy\cite{Brower1982,Lenahan2002,Poindexter1984}. With these techniques, average IT densities, and their combined effect on the global electronic properties of the device, are characterized. However, for both quantum and classical computation, it is increasingly important to understand how individual ITs modify the local electronic environment at the nanoscale.

The coherence of spin qubits in silicon, which are laterally spaced nanometers apart\cite{Wellard2003,GonzalezZalba2021} is determined in part by their electronic bath\cite{Wang2021,Ambal2016} (i.e. the silicon electronic landscape). So, nanoscale inhomogeneities of the silicon susceptibility (due to e.g. the presence of ITs) compromise qubit performance, which will become increasingly significant as qubit numbers continue to increase. Classical computation is also not immune to defect states at the Si surface: Indeed, as circuit components shrink to nanometer dimensions, surface and interface effects play an increasingly dominant role in device function\cite{Chen2019}. It is thus important to understand the origin of inhomogeneity in the electronic properties of nanoscale silicon-based devices.
Scanning probe microscopy (SPM) techniques allow for the characterization of trap densities and charge states in silicon with micron-scale\cite{Gramse2020,Izumi2023}, and nanoscale\cite{Ludeke2001,Johnson2009,Labidi2015,Turek2020} spatial resolution. However, trap charging and discharging timescales, which are non-zero and associated with energy loss under the influence of a time-varying electric field, have so far not been measured for individual ITs. In this work, we present the first spatially resolved measurements of dielectric loss associated with individual ITs at the Si/SiO\textsubscript{2} interface, characterized with nanometer spatial resolution using frequency-modulated atomic force microscopy (fm-AFM)\cite{Albrecht1991,Garcia2002}. We find that dielectric loss is spatially heterogeneous over nanometer length scales, and that the charge relaxation times at the Si/SiO\textsubscript{2} interface range between $1-150~\mathrm{ns}$, where there is a significant increase in this timescale around isolated trap states.

\begin{figure}[t]
    \centering
    \includegraphics[width=0.85\linewidth]{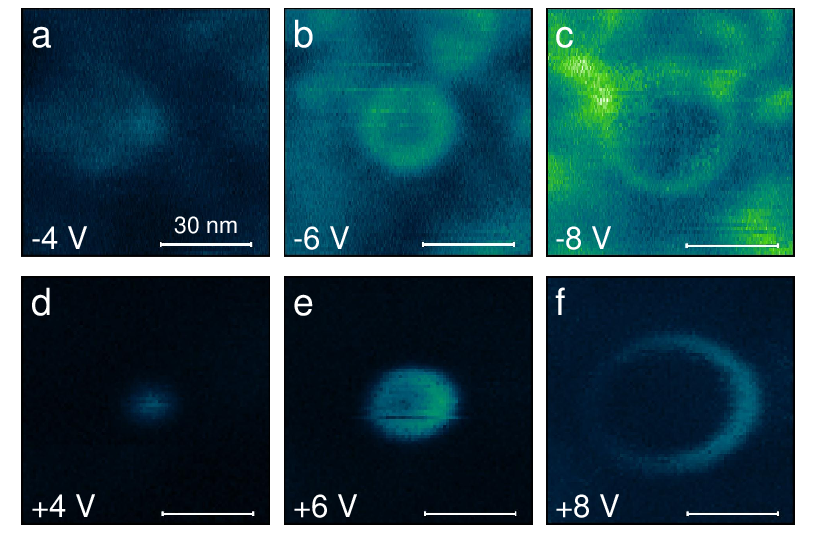}
    \caption{\textbf{ITs at the Si/SiO\textsubscript{2} interface}. fm-AFM $F_d$ measurement corresponding to dielectric loss at a donor-like (a-c) and acceptor-like (d-f) interfacial trap, measured at variable bias at room temperature in ultra-high vacuum (UHV). The colour scale is ${F_d=0~:550~\mathrm{meV/cycle}}$.}
    \label{fig:Rings}
\end{figure}

The sample studied here is a patterned n-type Si surface buried beneath $1~\mathrm{nm}$ of epitaxial silicon and terminated with $1~\mathrm{nm}$ of native oxide. Figure~\ref{fig:Rings} shows the spatial variability of the fm-AFM driving force $F_d$, which is an indirect measure of the dielectric loss, near two different trap sites observed at the Si/SiO\textsubscript{2} interface at variable tip-substrate gate bias $V_g$. The dielectric loss is bias-dependent and highly sensitive to spatially localized ITs. The measurement methodology will now be briefly discussed, before being applied to study the sample described above.

At large ($>10~\mathrm{nm}$) tip-sample separations in non-magnetic systems, where the tip-sample force is predominantly electrostatic, the fm-AFM tip-sample junction can be described as a metal-insulator-semiconductor (MIS) capacitor\cite{Cowie2022,Feenstra2006,Johnson2009,Johnson2011,DiBernardo2022,Wang2015,Winslow2011}. In this work, the MIS capacitor is comprised of a metallic tip, an insulating gap of thickness $z_{ins}$ (composed of the $~\sim 10~\mathrm{nm}$ tip-sample vacuum gap plus $1~\mathrm{nm}$ of SiO\textsubscript{2}), and an n-type Si(100) substrate. The total capacitance of this system is made up of the insulator (oxide and vacuum) and interfacial ($C_{int}$) capacitances in series, where $C_{int}$ describes the space-charge organization (i.e. band bending) at the silicon-oxide interface. For the low-frequency MIS capacitor, loss can be described as an equivalent series resistance (ESR) which introduces a phase shift in the circuit response corresponding to the Debye charging/discharging timescale $\tau$ of the Si/SiO\textsubscript{2} interface. In other words, $\tau$ is the time required to establish a surface potential $V_S$, which is non-zero due to the finite carrier mobility. 

fm-AFM is a dynamic microscopy, which is why it can be used to characterize dielectric loss\cite{Denk1991, Eslami2021,Oyabu2006,Suzuki2014,Kantorovich2004}. In fm-AFM, a cantilever-mounted tip is driven on the cantilever resonance $\omega$ at a constant oscillation amplitude $A$ above a sample surface. This means that over every oscillation cycle, the insulator thickness $z_{ins}$ varies in time. Consequently, the surface charge organization (i.e. band bending $V_S$) and tip-sample force ($\vec{F}_{ts}$\cite{Hudlet1995}) also vary in time. See the Supplementary Materials for animations of this dynamic sample response. $\vec{F}_{ts}$ leads to a shift in $\omega$ with respect to the free natural resonance $\omega_o$. Assuming harmonic oscillation where ${z_{ins}(t)=A\cos(\omega t)}$, the frequency shift $\Delta \omega$ and drive amplitude $F_d$ are\cite{Holscher2001,Sader2005,Kantorovich2004}:
\begin{subequations}\label{eq:dfdg}
\begin{eqnarray}
    \Delta\omega=\omega-\omega_o = \frac{-\omega_o}{2 kA}\frac{\omega_o}{\pi}\int_{0}^{2\pi/\omega}\partial t~F_{ts}(t)\cos(\omega t)~~~\\
    F_d = \frac{kA}{Q}-\frac{\omega_o}{\pi}\int_{0}^{2\pi/\omega}\partial t~F_{ts}(t)\sin(\omega t)~~~
\end{eqnarray}
\end{subequations}
\noindent where $k$ and $Q$ are the spring constant and Q-factor of the cantilever. In the derivation of Equation~\ref{eq:dfdg} (see \cite{Holscher2001,Kantorovich2004,Sader2005,Miyahara2015,Cowie2022}), Equation~\ref{eq:dfdg}b contains the phase information of the Fourier series expansion of $F_{ts}(t)$, such that $\Delta\omega$ is related to the components of $\vec{F}_{ts}(t)$ which are in-phase with $z_{ins}(t)$ and $F_d$ depends on the out-of-phase $\vec{F}_{ts}(t)$ components. A non-zero surface charge organization timescale $\tau$ therefore manifests as in increase in $F_d$\cite{Cowie2022}. fm-AFM, then, can be thought of as spatially localized admittance spectroscopy\cite{Nicollian1967,Engstrom2014,Piscator2009}, in which the MIS capacitance and conductance are measured by modulating the MIS potential. Typically in admittance spectroscopy, the MIS potential is modulated by applying an AC bias. In fm-AFM, the MIS potential modulation occurs inherently due to the oscillating cantilever.

The results shown in this work were measured in the low-frequency (quasi-static) regime (${f=2\pi\omega\approx 310~\mathrm{kHz}}$), such that an increase in $\tau$ means that more energy is dissipated by Ohmic loss\cite{Denk1991,Arai2018}. In other words, $\tau$ is the resistor-capacitor (RC) time constant of the MIS capacitor. In the small-angle regime where $\tau\ll1/f$, as in this experiment, the surface charge re-organization can be approximated as a constant phase offset $\delta$ between $F_{ts}(t)$ and $z_{ins}(t)$:
\begin{equation}\label{eq:delta}
    \delta = \tau \omega
\end{equation}
\noindent where $\omega=2\pi f$. 
\noindent Consequently, an increase in $\tau$ corresponds to an increase in the out-of-phase force component, and in the measured $F_d$. Since in the small-angle regime $\tan(\delta)\approx\delta$, Equation~\ref{eq:delta} shows that an increase in $\tau$ corresponds to an increase in the ESR or loss tangent $\tan(\delta)$.

In this work, $F_d$ is measured using fm-AFM to determine the spatial inhomogeneity of $\tau$. The experimental $\tau$ is calculated by comparing experimental $F_d$ bias spectra to modelled $F_d$ bias spectra at variable $\tau$. The modelled $F_d$ spectra are calculated by solving the MIS capacitor model as a function of time over an entire cantilever oscillation cycle (by varying the insulator thickness $z_{ins}$). See the Supplementary Materials for a description of the MIS model used in this work. The MIS force $F_{ts}(t)$ is then calculated and integrated according to Equation~\ref{eq:dfdg}, to solve for $\Delta f$ and $F_d$. This series of calculations is repeated for $\sim 10,000$ values of $\tau$. For each measured bias, the experimental $F_d$ is compared to all of the modelled results at that bias. The model which minimizes the difference between the experimental and modelled $F_d$ spectra is taken as the best-fit $\tau$. 

\begin{figure*}[t]
    \centering
    \includegraphics[width=\linewidth]{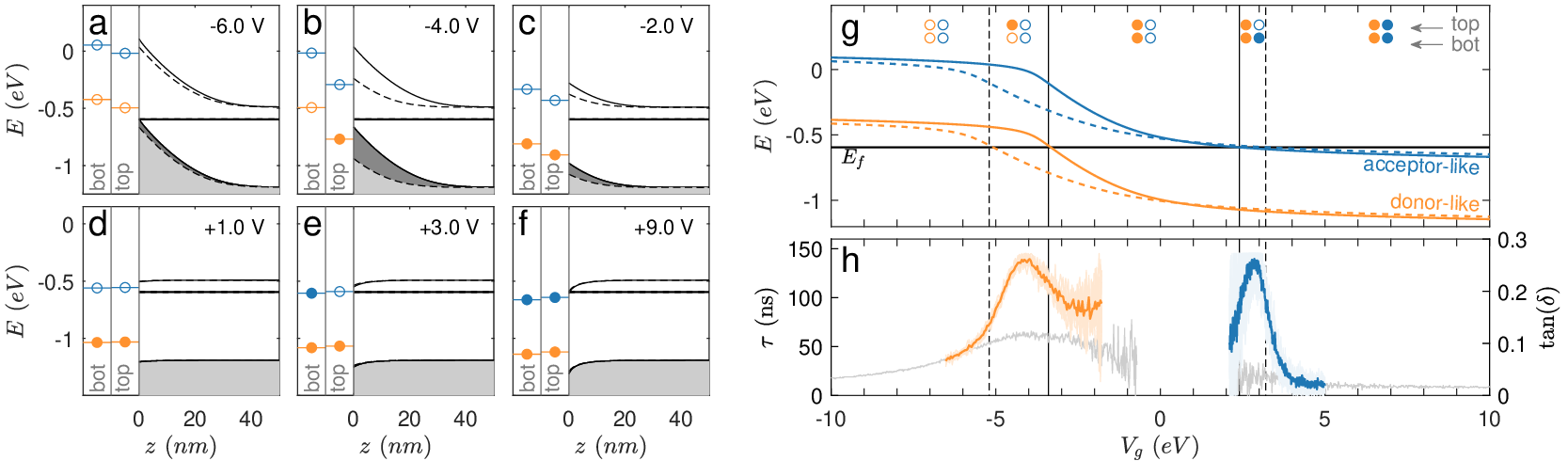}
     \caption{\textbf{Interfacial state occupancy and loss}. As the cantilever oscillates, the potential at the sample surface varies. (a-f) Modelled silicon band bending near the Si/SiO\textsubscript{2} interface ($z=0$) at the bottom (``bot", closest $z_{ins}$, solid) and top (``top", farthest $z_{ins}$, dashed) of the cantilever oscillation at different biases ($V_g$). (The dashed and solid curves nearly overlap at positive biases.) The tip and insulating gap are not shown.  Donor-like (orange) and acceptor-like (blue) states are shown at the bottom and top of the cantilever oscillation. The state occupancy is indicated by full (occupied) or empty (unoccupied) circles. (g) Modelled $V_g$-dependent energy of the donor-like and acceptor-like states relative to $E_f$ at the bottom (solid) and top (dashed) of the $6~\mathrm{nm}$ cantilever oscillation. The corresponding modelled crossing points, where the trap energy equals $E_f$, are indicated by vertical lines. (h) RC time constant $\tau$ and loss tangent $\tan(\delta)$ measured above a donor-like trap (orange), acceptor-like trap (blue), and far from either trap (grey). 10 curves are shown for each trap, with their average overlaid. In (h), the uncertainty diverges to infinity as $V_g$ approaches the flatband voltage $V_{fb}$, so this region is omitted.}
    \label{fig:StateOccupancy}
\end{figure*}

\section*{Results}
The sample measured here contains patterned squares of variable two-dimensional dopant density, up to a maximum of $1.6\mathrm\times 10^{14}/\mathrm{cm}^2$\cite{Stock2020}, on a background substrate doping of $9.0\times 10^{14}/\mathrm{cm}^3$. The un-patterned background is bulk doped with phosphorous, while the patterned squares are delta-doped with arsenic with a dopant layer thickness of approximately $2~\mathrm{nm}$. The entire wafer is capped by $3~\mathrm{nm}$ of epitaxial Si, the surface of which has subsequently formed $1~\mathrm{nm}$ of native SiO\textsubscript{2}, as determined by secondary mass ion spectroscopy\cite{Stock2020}. The results shown in Figures~\ref{fig:Rings} and \ref{fig:StateOccupancy} were measured in the background (lowest dopant density) region. 

\textbf{Spatial inhomogeneity--} The Si/SiO\textsubscript{2} interface is prone to trap states which modify the electronic properties of the MIS capacitor\cite{Kirton1989,Engstrom2014,Sze2007}. In particular, ITs (such as P\textsubscript{b0} and P\textsubscript{b1} centers) which have energy levels within the band gap interact significantly with the Si/SiO\textsubscript{2} interface charge\cite{Engstrom2014,Schroder2009}. Donor-like traps (e.g. Figure~\ref{fig:Rings}a-c), which have energies in the lower half of the band gap, can become positively charged via emission of an electron to the valence band. Acceptor-like traps (e.g. Figure~\ref{fig:Rings}d-f), which have energies in the upper half of the band gap, can become negatively charged via capture of an electron from the conduction band\cite{Sze2007,Schroder2009}. The interface state occupancy depends on $V_S$ (and therefore $V_g$), since capture or emission into a trap depends on its energy with respect to the Fermi level $E_f$\cite{Kirton1989}. This section shows that when the fm-AFM tip is positioned near an IT, the surface charging timescale $\tau$ increases. This is measured as an increase in the applied fm-AFM drive amplitude $F_d$.

Figure~\ref{fig:StateOccupancy}a-f show modelled band diagrams including a donor-like trap and an acceptor-like trap at the bottom and top of the fm-AFM cantilever oscillation. The donor-like trap is unoccupied at high negative voltage, but as $|V_g|$ decreases and the bands flatten and bend downward, the donor-like trap energy lowers below $E_f$, and it becomes occupied. The acceptor-like trap is unoccupied from negative biases up to positive biases, where the trap energy lowers below $E_f$ and it becomes occupied. The trap state energies found here ($0.17~\mathrm{eV}$ above the valence band for the donor-like trap and $0.65~\mathrm{eV}$ above the valence band for the acceptor-like trap) are in agreement with accepted levels for $P_{b0}$ states\cite{Poindexter1984,Ragnarsson2000}. Animations of the IT charge states as the cantilever oscillates are shown for variable bias $V_g$ in the Supplementary Materials.

Figure~\ref{fig:StateOccupancy}g shows the bias-dependent energy of each trap. At biases between the crossing points, the trap energy shifts above and below $E_f$ during every oscillation cycle (Figures~\ref{fig:StateOccupancy}b,e) and there is a significant increase in $\tan(\delta)$ as compared to the trap-free spectrum (Figure~\ref{fig:StateOccupancy}h). This is because, upon electron capture and re-emission as the cantilever oscillates, loss occurs as the system relaxes to its ground state in a mechanism attributed to cascade phonon scattering\cite{Nicollian1967,Lax1960,Wang2007,Piscator2009}. The magnitude of $\tan(\delta)$ measured here is consistent with previously reported bias-dependent loss peaks attributed to interface states in silicon\cite{Yldz2011,BirkanSeluk2007}. The grey spectrum in Figure~\ref{fig:StateOccupancy}h corresponds to the intrinsic relaxation timescale of the Si/SiO\textsubscript{2} interface. The orange and blue spectra show the increase in the surface relaxation timescale near Si/SiO\textsubscript{2} ITs. Note that the grey spectrum in Figure~\ref{fig:StateOccupancy}h is non-zero and also varies with bias. An explanation of the origin of this background bias dependence (i.e. why $\tau$ is non-zero even in the absence of ITs) will follow.

\begin{figure}[t]
    \centering
    \includegraphics[width=0.83\linewidth]{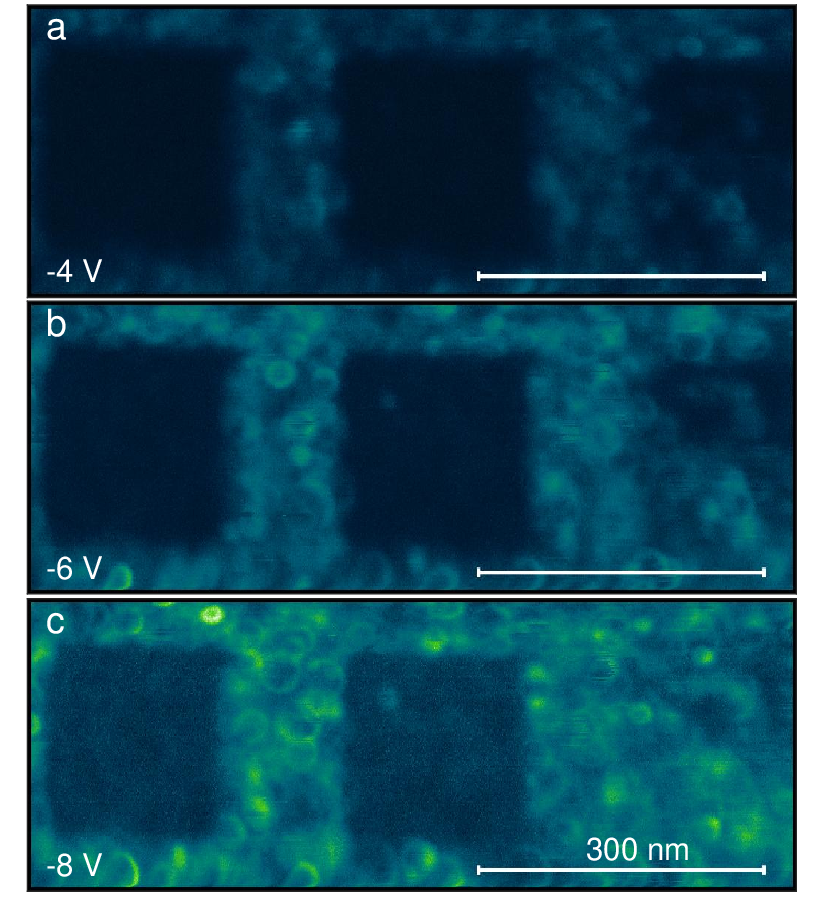}
    \caption{\textbf{Spatial inhomogeneity of Si/SiO\textsubscript{2}.} fm-AFM $F_d$ measurement of the patterned surface. The dopant density is highest in the two left-most squares; intermediate in the right square; and lowest in the background. The colour scale bar is ${F_d=0:500~\mathrm{meV/cycle}}$.}
    \label{fig:Squares}
\end{figure}

The bias-dependent spatial inhomogeneity of $\tan(\delta)$ manifests as the ring-like $F_d$ features in Figures~\ref{fig:Rings} and \ref{fig:Squares}\cite{Turek2020}. Any spatially localized process which exhibits a peak in a bias spectrum in fm-AFM manifests as a ring when imaged spatially at constant height\cite{Cockins2010}, due to the spatial localization of the top gate (tip), which introduces circularly symmetric equipotential lines at the sample surface. As the tip moves in $x$, $y$, or $z$ away from a trap, the peak shifts to more extreme biases. This is demonstrated in the Supplementary Materials. 

\textbf{Dopant density dependence--} Three ``delta-doped" patterned squares of this sample can be seen in Figure~\ref{fig:Squares}. Almost no rings appear in the highly doped (square-patterned) regions. This is due to the increased Si metallicity within the the patterned squares: As the dopant density increases, the change in band bending over every fm-AFM oscillation cycle ($\Delta V_S$) decreases, meaning that defects are not electronically accessed according to the process shown in Figures~\ref{fig:StateOccupancy}a-f (that is, the crossing points would occur at $V_g<-10~V$).  By similar reasoning, acceptor-like rings are much sparser: Measurements at positive bias -- not shown -- exhibited fewer than 5 rings over the area shown in Figure~\ref{fig:Squares}. This is because at positive biases (in the accumulation regime), $\Delta V_S$ is small as the cantilever oscillates, so only states very close to the conduction band edge have crossing points at $V_g<10~V$. The donor-like IT density in Figure~\ref{fig:Squares} is approximately $10~\mathrm{traps}/100~nm^2$.


\begin{figure}[b]
    \includegraphics[width=\linewidth]{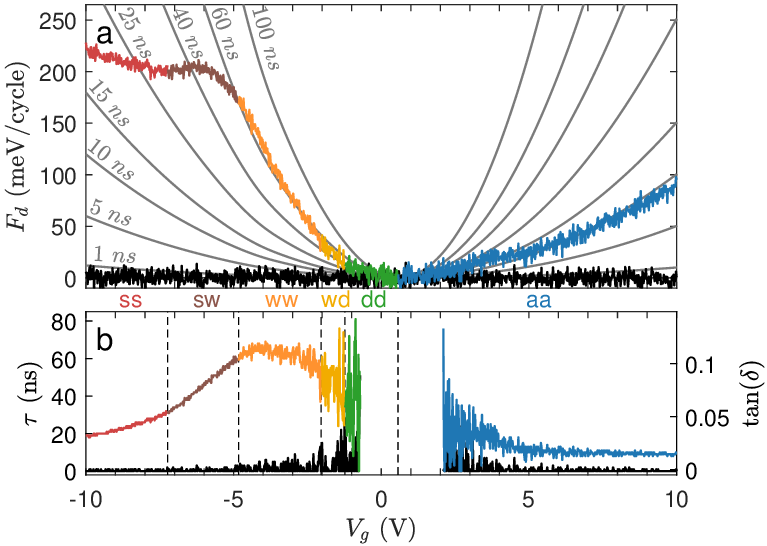}
    \caption{\textbf{Bias dependencies of an MIS capacitor}. (a) Measured drive amplitude $F_d$ with the tip close to the surface (color) and far from the surface (black, $z_{ins}\sim1~\mu m$). Modelled $F_d$ for various $\tau$ (indicated) are also shown. Six regimes (ss, sw, ww, wd, dd, and aa) are identified, indicating the bias regime (s --- strong inversion, w -- weak inversion, d -- depletion, or a -- accumulation) at the bottom and top of the oscillation. (For example, in the wd regime, the capacitor is under weak inversion at the closest tip-sample separation, and depletion at the farthest tip-sample separation). (b) Corresponding RC time constant $\tau$ and loss tangent $\tan(\delta)$. At the flatband voltage $V_{fb}$, indicated by a dashed line, the $F_d$ curves all overlap and the $\tau$ and $\tan(\delta)$ uncertainties diverge to infinity, so this region was omitted.}
    \label{fig:BiasDependence}
\end{figure}

\textbf{Bias dependence--} Even in the absence of ITs, the surface charge density continually re-organizes over every cantilever oscillation cycle. The nature of the surface charge re-organization is bias-dependent, and can be understood in terms of the bias regimes of the MIS capacitor at the bottom and top of the cantilever oscillation, as defined in the Figure~\ref{fig:BiasDependence} caption. Animations of the MIS bias regimes as the cantilever oscillates is shown in the Supplementary Materials. 

The dielectric loss measured here depends in part on the number of carriers moving within the depletion region as the cantilever oscillates. (See the Supplementary Materials for a more detailed explanation.) When the cantilever oscillates within the depletion and weak inversion regimes, $\Delta V_S$ is large over every oscillation cycle. Correspondingly, the number of holes moving within the depletion region as the cantilever oscillates is large, so $\tau$ and $\tan(\delta)$ (Figure~\ref{fig:BiasDependence}b) are large. When the cantilever oscillates within the strong inversion and accumulation regimes, $\Delta V_S$ is small and the number of carriers moving as the cantilever oscillates is small, so $\tau$ and $\tan(\delta)$ decrease. This signifies that dielectric loss corresponding to the surface charge re-organization in response to a time-varying MIS potential is inherently bias-dependent.

\section*{Conclusions}

We show that the magnitude of dielectric loss at the Si/SiO\textsubscript{2} interface is highly inhomogeneous\cite{Ambal2016,Turek2020,Winslow2011}, dopant density-dependent, and gate bias-dependent. In particular, ITs lead to a dramatic increase in dielectric loss at biases corresponding to the trap state energy. This result is directly applicable to fixed-geometry field-effect devices. In such devices, where the MIS potential is modulated by applying an AC gate bias, increasing the AC bias amplitude will increase the width of this bias-dependent loss peak. Increasing the distance between the gate and the trap (or, if a qubit is acting as a spectrometer of the trap state\cite{Wang2021}, the distance between the qubit and the trap) will shift the peak toward more extreme voltages. 

The surface charge organization timescale $\tau$ measured here is spatially variable and ranges between ${1-150~\mathrm{ns}}$, which encompasses typical Rabi frequencies of buried spin qubits (which are between $1-10~\mathrm{MHz}$)\cite{Huang2019,Kawakami2014}. This indicates that the amplitude and phase of the potential at qubit locations will be a function of the temporal structure of the applied bias pulse sequence and the local position and energy level of defect states. 

The values of $V_g$ in this work are much greater than the typical $\mu V-\mathrm{mV}$ values used for spin qubit readout. However, the bias-dependent MIS surface potential is highly sensitive to the capacitor geometry, specifically the insulator thickness. Here, the closest tip-sample separation is $12~\mathrm{nm}$, but for the same capacitor with a $1~\mathrm{nm}$ insulator thickness, the peak which occurs at $\sim -4~\mathrm{V}$ in Figure~\ref{fig:StateOccupancy}h can be expected to occur closer to $-500~\mathrm{mV}$. 

Finally, the $\tan(\delta)\sim 0.1$ measured here (which is similar to other room temperature findings\cite{Yang2006,Krupka2015,Krupka2006,Mukherjee2016}) is several orders of magnitude smaller at cryogenic temperatures\cite{Checchin2022,Liu2006}, as carrier concentrations decrease and various phonon scattering mechanisms are reduced\cite{Mukherjee2016}. Still, these dielectric losses can occur under any time-varying electric field, and so should be taken into consideration for the continued development of nano- and atomic-scale semiconductor devices, quantum sensors, and quantum computers. 


\section*{Methods}
\textbf{Experimental setup--} Nanosensors platinum-iridium coated silicon tips (PPP-NCHPt) with $\sim 310~\mathrm{kHz}$ resonant frequency, spring constant $42~\mathrm{N/m}$, and a Q-factor of approximately $18000$ were used for all measurements. The oscillation amplitude was $6~\mathrm{nm}$. Experiments were conducted at room temperature (assumed to be $300~\mathrm{k}$) in ultra-high vacuum ($\sim 10^{-10}~\mathrm{mbar}$). 

\textbf{Bias spectroscopy--} Each bias spectrum includes the forward (positive to negative $V_g$) and backward curve superimposed, showing that there is negligible hysteresis with bias. Each sweep was acquired over $\sim30~s$. 

\textbf{Multipass imaging--} Figures~\ref{fig:Rings} and \ref{fig:Squares} were measured by electrostatic force microscopy (EFM) multi-pass imaging. In the first pass, the tip tracked the topography defined by $V_g=0~\mathrm{V}$ at a setpoint $\Delta f=-3~\mathrm{Hz}$. In subsequent passes, the tip followed this same topography, but $V_g$ was set to the displayed values. The rings shown here were stable over several weeks of measurement. 

\textbf{Sample fabrication--} The Si(001) substrate is phosphorous-doped ($9.15\times10^{14}\mathrm{/cm^3}$) and $300~\mu m$ thick. The variably arsenic delta-doped regions were fabricated by hydrogen resist lithography\cite{Stock2020}.

\textbf{MIS model--} The MIS model\cite{Sze2007,Hudlet1995,Cowie2022} parameters were: Closest $z_{ins}=12~\mathrm{nm}$, tip radius $5~\mathrm{nm}$, $\epsilon=11.7$, electron affinity $4.05~\mathrm{eV}$, tip work function $4.75~\mathrm{eV}$, electron and hole effective masses $1.08$ and $0.56$, n-type dopant density $5\times10^{17}\mathrm{/cm^3}$, and band gap $0.7~\mathrm{eV}$. This band gap is smaller than the $\sim 1.1~\mathrm{eV}$ expected for bulk Si; the discrepancy could be due to surface band gap narrowing due to the presence of the large surface state density, as in \cite{vanVliet1986,King2010}. For details of the MIS model, see the Supplementary Materials.

\begin{acknowledgments}
This research was supported by Natural Sciences and Engineering Research Council of Canada (NSERC) Alliance Grants -- Canada-UK Quantum Technologies, an NSERC Discovery Grant, and Fonds de recherche du Qu\'{e}bec -- Nature et technologies, as well as the Engineering and Physical Sciences Research Council [grants EP/R034540/1, EP/V027700/1, and EP/W000520/1] and Innovate UK [grant 75574]. The authors would also like to thank Kirk Bevan and Hong Guo for stimulating discussions. 
\end{acknowledgments}


\bibliography{mainbib}


\newpage
\onecolumngrid
\setstretch{1.15}
\pagenumbering{gobble}

\newpage
\begin{center}
\huge Supplementary Materials
\end{center}
\begin{center}
\large for\\
\vspace{6mm}
\large{\textbf{Spatially resolved dielectric loss at the Si/SiO\textsubscript{2} interface}}\\
\vspace{2mm}
\normalsize{Megan Cowie\textsuperscript{1}, Taylor J.Z. Stock\textsuperscript{2,3}, Procopios C. Constantinou\textsuperscript{2}, Neil J. Curson\textsuperscript{2,3}, and Peter Gr\"{u}tter\textsuperscript{1}}\\
\vspace{1mm}
\small{\textsuperscript{1}\textit{Department of Physics, McGill University, Montr\'eal, Qu\'ebec, Canada}}\\
\small{\textsuperscript{2}\textit{London Centre for Nanotechnology, University College London, London, United Kingdom}}\\
\small{\textsuperscript{3}\textit{Department of Electronic and Electrical Engineering, \\University College London, London, United Kingdom}}
\end{center}
\bigskip 

This supplementary materials section contains five sections, listed below.\\

\hspace{10mm} Section I. MIS model description (GIFs 1-2)

\hspace{10mm} Section II. MIS fm-AFM description (GIFs 3-4)

\hspace{10mm} Section III. Conductivity calculations

\hspace{10mm} Section IV. Peak shape dependencies

\hspace{10mm} Section V. Model fit and parameter sensitivities\\

\newpage
\section{MIS model description} \label{Supp:MISmodeldescription}
\vspace{-4mm}
This section provides more detail about the modelled results shown in this work, which were calculated by describing the tip-insulator-sample junction as a one-dimensional metal-insulator-semiconductor (MIS) capacitor. In the one-dimensional MIS model\cite{Sze2007}, the $z$-dependent charge density $\rho_z(z)$ near the semiconductor surface is:
\begin{equation}\label{eq:rho_z}
    \rho_z(z)=|e|\left(p_z(z)-n_z(z)-N_a+N_d\right)
\end{equation}
\noindent where $e$ is the fundamental charge, $n_z(z)$ and $p_z(z)$ are the $z$-dependent electron and hole densities, respectively, and $N_d$ and $N_a$ are the ionized donor-like and acceptor-like dopant densities. Given the relations: 
\begin{subequations}\label{eq:n_z}
\begin{alignat}{2}
    n_z(z)=n~exp\left(\frac{|e|V_z(z)}{k_B T}\right)\\
    p_z(z)=p~exp\left(\frac{-|e|V_z(z)}{k_B T}\right)
\end{alignat}
\end{subequations}
\noindent where $n$ and $p$ are the bulk free electron and hole densities, respectively, the $z$-dependent electric field $\mathcal{E}_z(z)$ can be found given the one-dimensional Poisson equation as: 
\begin{equation}\label{eq:E_MIS}
    \mathcal{E}_z^2(z)=\frac{k_B T}{\epsilon}\left[
    p\left(exp\left(\frac{-|e|V_z(z)}{k_B T}\right)
    +\frac{|e|V_z(z)}{k_B T}-1\right)+ n\left(exp\left(\frac{|e|V_z(z)}{k_B T}\right)-\frac{|e|V_z(z)}{k_B T}-1\right) \right]
\end{equation}
\noindent where $k_B$ is the Boltzmann constant and $T$ is the temperature. In a net charge-neutral system: \begin{equation}\label{eq:neutrality}
0 = p-n+N_d-N_a
\end{equation}
\noindent The bulk Fermi level $E_f$ can be calculated numerically given the bulk definitions for $n$ and $p$:
\begin{subequations}\label{eq:np_effectiveDOS}
\begin{alignat}{2}
    n &= N_c\:\exp\left(\frac{E_f-E_c}{k_BT}\right)\\
    p &= N_v\:\exp\left(\frac{E_v-E_f}{k_BT}\right)
\end{alignat}
\end{subequations}
\noindent where $N_c$ and $N_v$ are the effective number of conduction and valence band states (defined in e.g. \cite{Sze2007}) and $E_c$ and $E_v$ are energies of the conduction and valence band edges in the bulk. The surface potential $V_S=V_z(z=0)$ is calculated as the numerical solution to the potential continuity equation of the MIS capacitor: 
\begin{equation}\label{eq:potcontinuity_MIS}
    0=V_g-V_{CPD}-V_S-V_{ins}
\end{equation}
\noindent where $V_g$ is the applied gate voltage between the tip and sample, $V_{CPD}$ is the contact potential difference between the metal and the semiconductor, and $V_{ins}=-eQ_S/C_{ins}$ is the potential drop across the insulator. For the model used in this work, the insulating gap (which experimentally is comprised of $1~\mathrm{nm}$ SiO\textsubscript{2} and $\sim 10-20~\mathrm{nm}$ vacuum which varies as the cantilever oscillates) is approximated to be entirely comprised of vacuum, such that $C_{ins}=\epsilon_o/z_{ins}$. The full $z$-dependent band diagram (i.e. band bending) is calculated according to:
\begin{equation}
    z'=\int_{V_S}^{V_{z}'} \frac{1}{\mathcal{E}(V)}\partial V
\end{equation}

The force between the MIS capacitor plates (i.e. the force between the metallic tip and the semiconducting sample) is predominantly electrostatic at the large ($\sim 10-20~\mathrm{nm}$) tip-sample separations used in this work\cite{Hudlet1995}: 
\begin{equation}\label{eq:F_MIS}
    F_{ts} = \frac{-Q_S^2}{2\epsilon_o}
\end{equation}
\noindent where $Q_S=Q_z(z=0)$), which is calculated from the MIS capacitor model above.

\section{MIS fm-AFM description} \label{Supp:MISfmAFMdescription}
\begin{figure}[!h]
    \centering
    \includegraphics[width=0.5\linewidth]{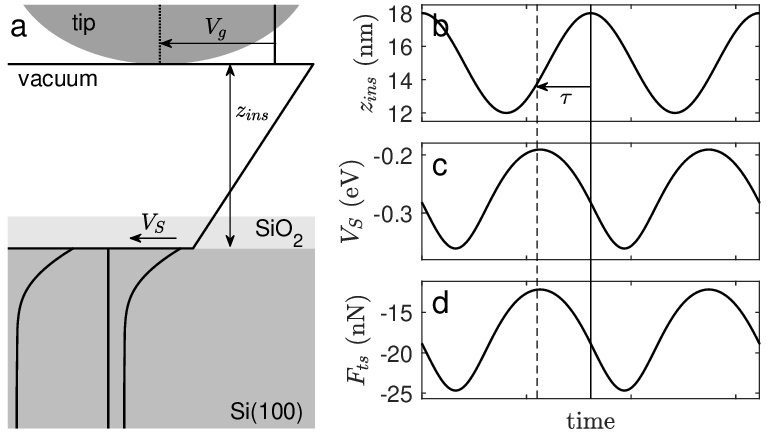}
    \caption{\textbf{Experimental setup.} (a) fm-AFM tip-sample junction with a modelled MIS capacitor band diagram (not drawn to scale) overlaid. The gate bias $V_g$ is applied to the tip and the sample is grounded. (b-d) Insulator thickness $z_{ins}$, surface potential $V_S$, and tip-sample force $\vec{F}_{ts}$ are shown for $V_g=-3~V$ over two cantilever oscillations. The RC time constant $\tau$ is exaggerated by several orders of magnitude for illustrative purposes. (The $\tau<150~ns$ measured in this work would be indistinguishable here if drawn to scale.)}
    \label{fig:ExperimentalSetup}
\end{figure}

\vspace{-4mm}
This section elaborates on the how the MIS model is applied to an oscillating fm-AFM cantilever. In a fm-AFM experiment, the insulator thickness changes as the cantilever oscillates. Consequently (given Equation~\ref{eq:potcontinuity_MIS} above) $V_S$ varies in time. Figure~\ref{fig:ExperimentalSetup} illustrates the fm-AFM tip-sample junction, and shows modelled results showing the time-dependent insulator thickness $z_{ins}(t)$, and correspondingly time-dependent surface potential $V_S(z(t))$ and tip-sample force $F_{ts}(z(t))$. $V_S$ also depends on the gate bias $V_g$ according to Equation~\ref{eq:potcontinuity_MIS}. Animated MIS band diagrams showing the time-dependent band band bending at variable $V_g$, calculated according to the MIS model above and the parameters defined in the main text, are shown in GIFs~1 and 2. 

In this work, the cantilever period is approximately $3~\mathrm{\mu s}$ (corresponding to its $f\sim 300~\mathrm{kHz}$ resonance frequency) and $\tau$, the charge re-organization timescale (effectively the RC time constant or Debye charging timescale of the Si/SiO\textsubscript{2} interface) is $\sim 100~\mathrm{ns}$. In this regime, where $\tau\ll1/f$, the capacitor can be treated as quasi-static throughout the cantilever oscillation, and the MIS model outlined above applies. Animated diagrams showing the modelled relationship between $z_{ins}$, $V_S$, $F_{ts}$, as well as the modelled charge organization in the MIS capacitor as the cantilever oscillates, for both $\tau=0$ and $\tau>0$, are shown in GIF~3. This effect can also be understood in phasor representation; animated phasor diagrams are shown in GIF~4.

Note that the relationship between the fm-AFM drive $F_d$ and the gate bias $V_g$ for any $\tau$ is highly non-linear, as can be seen in Figure~\ref{fig:BiasDependence}a of the main text. Therefore, the $F_d$ imaging contrast (see e.g. Figures~\ref{fig:Rings} and \ref{fig:Squares}) is also bias-dependent. At large negative $V_g$, for example, the measured $F_d$ increases significantly for a given increase in $\tau$, whereas at small negative $V_g$, the same increase in $\tau$ corresponds to a smaller increase in $F_d$. Consequently, for example, the rings in Figure~\ref{fig:Squares}c at $V_g=-8~\mathrm{V}$ appear brighter than the dots in Figure~\ref{fig:Squares}a at $V_g=-4~\mathrm{V}$, even though they correspond to the same $\tau$. 

\newpage
\section{Conductivity calculation} \label{Supp:Conductivity calculation}
\begin{figure}[!h]
    \centering
    \includegraphics[width=\linewidth]{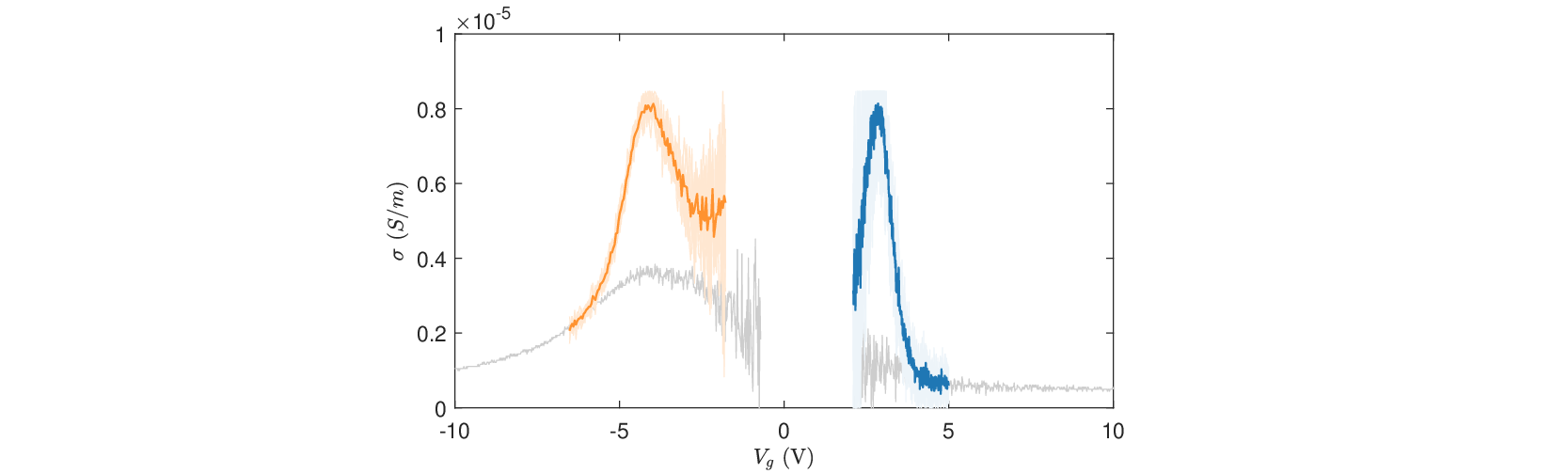}
    \caption{\textbf{Bias-dependent conductivity.} Conductivity $\sigma$ calculated according to Equation~\ref{eq:losstangentconductivity}. The colours correspond to 
Figure~\ref{fig:StateOccupancy} in the main text (grey is measured with the tip far from any trap; light orange is repeated measurements above a donor-like IT, with their mean shown as dark orange; light blue is repeated measurements above an acceptor-like IT, with their mean shown as dark blue).}
    \label{fig:BiasDependentConductivity}
\end{figure}

\vspace{-4mm}
This section shows the AC conductivities $\sigma$ calculated given the experimental loss tangents $\tan(\delta)$ measured in this work. The finite charge re-organization timescale $\tau$ corresponds to energy dissipation in the sample as the cantilever oscillates, where in the small-angle limit (as in this experiment), the loss tangent $\tan(\delta)\approx\delta$.  In this low-frequency regime, loss is dominated by scattering of free carriers, which is related to the sample conductivity $\sigma$ according to\cite{Checchin2022,Krupka2006,Krupka2015}:  
\begin{equation}\label{eq:losstangentconductivity}
    \tan(\delta)=\frac{\sigma}{\omega\epsilon}
\end{equation} 
\noindent An estimation of $\sigma$, given Figure~\ref{fig:BiasDependence} in the main text and Equation~\ref{eq:losstangentconductivity}, is shown in Figure~\ref{fig:BiasDependentConductivity}.

$\sigma$ is related to the mobility $\mu_{tot}$, as well as $N$, the number of carriers moving in the depletion region as the cantilever oscillates, as:
\begin{equation}\label{eq:mobility}
    \sigma = \mu_{tot}Ne 
\end{equation}
\noindent Therefore, $\tau$ and $\tan(\delta)$ as measured in this work can be understood in the context of the free carrier mobilities, which depend on the various (spatially variable) scattering timescales of the material. This spatial dependence, specifically the elongation of $\tau$ and corresponding increase in $\tan(
\delta)$ near ITs, is the focus of Figure~\ref{fig:StateOccupancy} in the main text.

Additionally, Equation~\ref{eq:mobility} indicates that $\tau$ and $\tan(\delta)$ depend on $N$. GIF~1 shows that far from any ITs, $N$ is highly bias-dependent. This is the origin of the bias-dependent $F_d$ for the background spectrum shown in Figures~\ref{fig:StateOccupancy}h, and is the focus of Figure~\ref{fig:BiasDependence} in the main text. In other words, even for a constant $\mu_{tot}$, $F_d$ is expected to be asymmetric in bias.

\newpage
\section{Peak shape dependencies} \label{Supp:PeakShapeDependencies}
\vspace{-4mm}
This section demonstrates the experimental dependencies of the fm-AFM drive $F_d$ and frequency shift $\Delta f$ on the oscillation amplitude $A$, tip-sample separation $z$, and gate voltage $V_g$.

\begin{figure*}[b]
    \centering
    \includegraphics[width=0.85\linewidth]{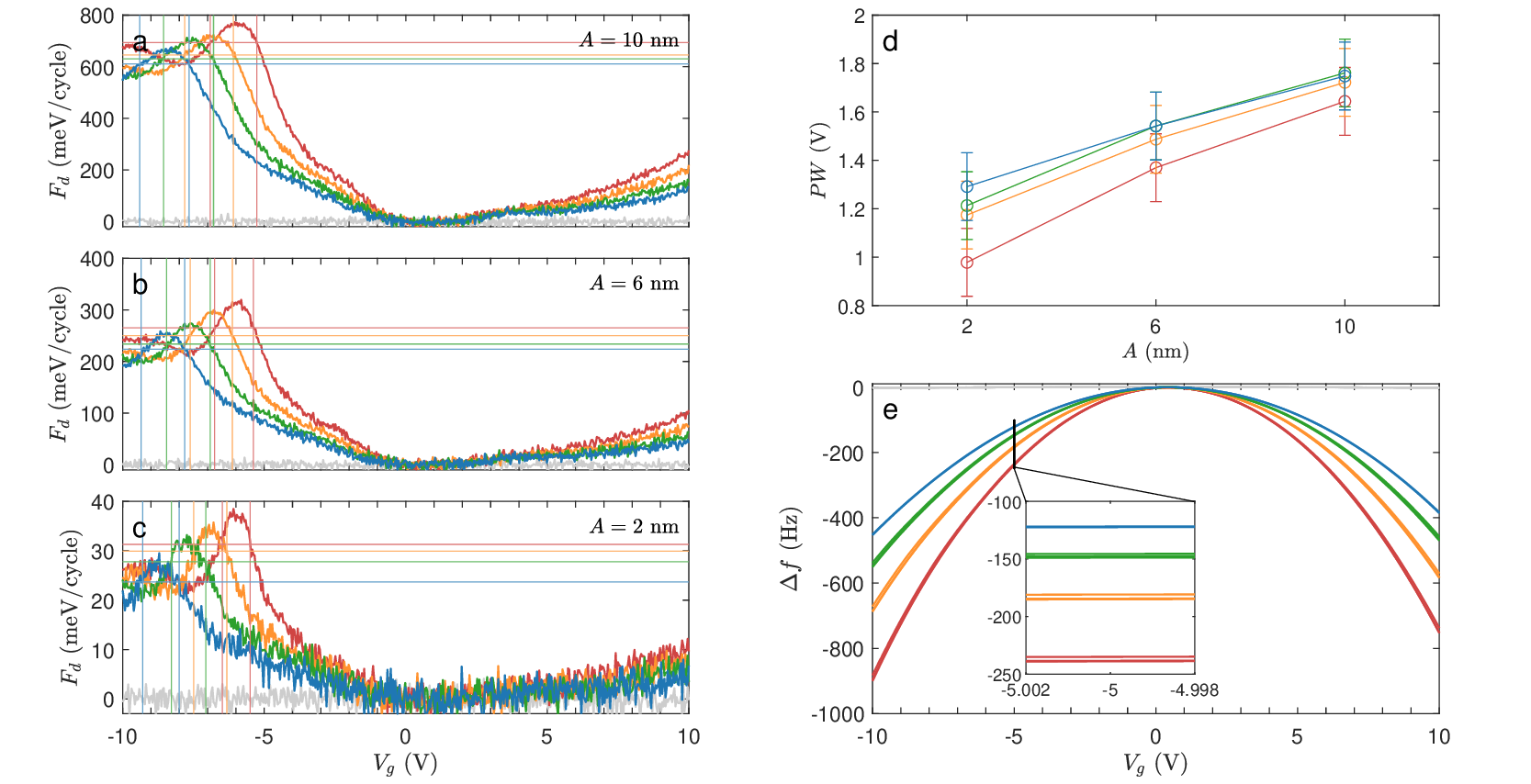}
     \caption{\textbf{Bias spectra: Oscillation amplitude ($A$) and tip-sample separation ($z$) dependencies}. (a-c) Drive amplitude $F_d$ bias spectra measured for $A=10$, $6$, and $2~\mathrm{nm}$ at variable tip lift $z_{TL}$ ($z_{TL}=0~\mathrm{nm}$, red; $z_{TL}=2~\mathrm{nm}$, yellow; $z_{TL}=4~\mathrm{nm}$, green; $z_{TL}=6~\mathrm{nm}$, blue). (d) $A$ dependence of the $F_d$ peak width ($PW$). The relationship is expected to be direct but not linear, and the lines shown are not fits. (e) $\Delta f$ bias spectra measured simultaneously with the $F_d$ spectra in (a-c). All 15 bias spectra are shown in (e), but the fm-AFM frequency shifts $\Delta f$ for variable $A$ are essentially overlapping at each $z_{TL}$. }
    \label{fig:PeakShapeDependencies}
\end{figure*}

The positions of the peaks shown in Figure~3 of the main text depend on the DC bias at the trap state position. Therefore, the peak position depends on the distance between the tip (localized gate) and the IT. When the distance is increased, the potential experienced at the trap site is reduced, and so the peak position shifts to more extreme values. This is why these peaks manifest as the rings shown in Figures~1 and 4 of the main text. Here, we show the $z$ distance dependence on the peak position, at a slight lateral $x,y$ displacement from an n-type ring center.  Figure~\ref{fig:PeakShapeDependencies}a shows bias spectra at variable tip lift ($z_{TL}$) above a $-3~\mathrm{Hz}$ setpoint frequency with a $10~\mathrm{nm}$ oscillation amplitude $A$. At the smallest $z_{TL}$ (red), the peak occurs at $V_g\sim -6~\mathrm{V}$; at the largest $z_{TL}$ (blue), the peak occurs at $V_g\sim -8.5~\mathrm{V}$; therefore, a $6~\mathrm{nm}$ increase in the distance between the trap site and the tip corresponds to a $\sim 2.5~\mathrm{V}$ peak shift. This result agrees with the rings shown in Figure~1 in the main text, which shows that the ring radius increases by $\sim 5~\mathrm{nm}$ when $V_g$ increases by $2~\mathrm{V}$. 

The peak shape also depends on the amplitude of the MIS bias modulation, and for this fm-AFM experiment is therefore related to the tip oscillation amplitude $A$. Figures~\ref{fig:PeakShapeDependencies}a-c show bias spectra measured at $A=10$, $6$, and $2~\mathrm{nm}$. As $A$ increases, the peak width ($PW$) increases. The direct correlation between $A$ and $PW$ is demonstrated in Figure~\ref{fig:PeakShapeDependencies}d. $PW$ was found by measuring $V_{g,L}$, the bias corresponding to the half-maximum $F_d$ value on the left side of the peak; $V_{g,R}$ is the bias value corresponding to this $F_d$ on the right side of the peak. The total peak width is their difference: $PW=V_{g,R}-V_{g,L}$. The $V_{g,L}$ and $V_{g,R}$ values are the vertical lines in Figures~\ref{fig:PeakShapeDependencies}a-c; the half-maximum $F_d$ values are the horizontal lines. The uncertainty in $A$ is negligible on these axes. The uncertainty in $PW$ is dominated by the estimation of the half-maximum $V_g$ on both sides of the peak. Approximating these uncertainties each as $\sim 100~\mathrm{mV}$ gives a peak width uncertainty of $0.14~\mathrm{V}$.

Figure~\ref{fig:PeakShapeDependencies}e shows the $\Delta f$ bias spectra measured simultaneously with the $F_d$ spectra in Figures~\ref{fig:PeakShapeDependencies}a-c. $\Delta f$, which corresponds to the in-phase components of $F_{ts}$ according to Equation~2a in the main text, inversely depends on $z_{TL}$: As $z_{TL}$ increases, band bending at the semiconductor surface decreases, meaning that $F_{ts}$ is reduced. $\Delta f$ is approximately independent of $A$. This is demonstrated in the inset of Figure~\ref{fig:PeakShapeDependencies}e, which shows that the $A=10$, $6$, and $2~\mathrm{nm}$ $\Delta f$ curves are essentially overlapping at each $z_{TL}$.

\newpage
\section{Model fit and parameter sensitivities} \label{Supp:FitParameterSensitivies}
\vspace{-4mm}
This section demonstrates that the fm-AFM frequency shift $\Delta f$ and drive amplitude $F_d$ bias spectra are experimentally reproducible. Additionally, this section demonstrates that the modelled MIS $\Delta f$ bias spectra are in good agreement with the experimental data. The MIS model parameter sensitivities are also shown. 

In this work, the in-phase force (related to $\Delta f$) and the out-of-phase force (related to $F_d$, see Equation~1 in the main text) are analyzed to determine $\tau$. First, the MIS model is fitted to the $\Delta f$ spectra which, given that $\tau$ is small with respect to the cantilever oscillation period, is independent of $\tau$ within the measurement sensitivity. Modelled $F_d$ spectra at variable $\tau$ are then compared to the experimental $F_d$ results to give the best fit $\tau$ value. 

\begin{figure}[b]
    \centering
    \includegraphics[width=\textwidth]{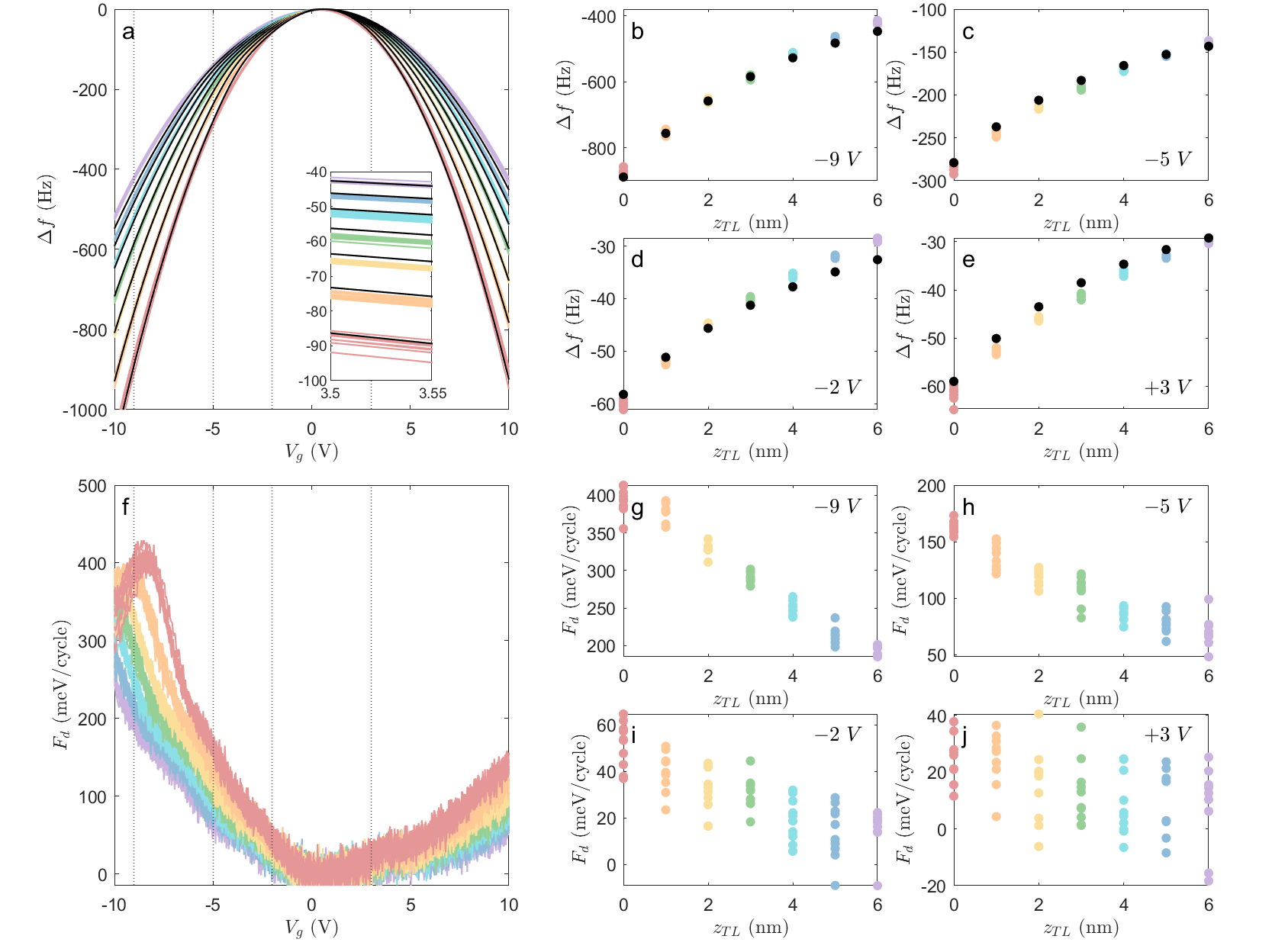}
    \includegraphics[width=\textwidth]{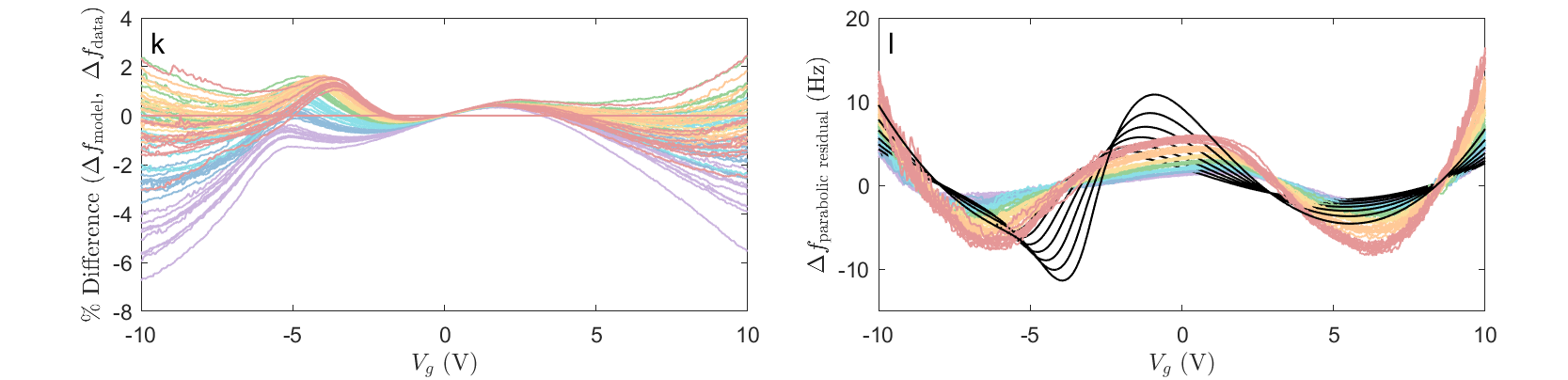}
    \caption{\textbf{Comparison of Si measurements and the MIS model.} Experimental bias spectra (colours) for variable tip lift: $z_{TL}=0~\mathrm{nm}$ (red), $1~\mathrm{nm}$ (orange), $2~\mathrm{nm}$ (yellow), $3~\mathrm{nm}$ (green), $4~\mathrm{nm}$ (teal), $5~\mathrm{nm}$ (blue), and $6~\mathrm{nm}$ (purple).  For each $z_{TL}$, ten bias spectra are shown. (a) shows the fm-AFM frequency shift $\Delta f$, and (c-e) show $\Delta f$ at the $V_g$ indicated, with the model (black) overlaid. (f-j) show the fm-AFM drive amplitude $F_d$.  (k) shows the percent difference between the $\Delta f$ data and the model for each spectrum. (l) shows the difference between each spectrum (and model) with its parabolic fit, which demonstrates the similar non-parabolicity of both the data and the model.} 
    \label{fig:Supp_TipLiftComparison}
\end{figure}

Figure~\ref{fig:Supp_TipLiftComparison} demonstrates the quality of the MIS model fit to the experimental bias spectra. Repeated measurements of $\Delta f$ and $F_d$ were performed at variable tip lift ($z_{TL}$) slightly displaced from a ring center (a,f). The repeated curves overlap, which shows that the measurements are robust. $\Delta f$ and $F_d$ are also shown for four arbitrary biases ($V_g=-9\mathrm{V}$, $-5~\mathrm{V}$, $-2~\mathrm{V}$, and $+3~\mathrm{V}$ as a function of $z_{TL}$ (b-e, g-j). The MIS model fit for variable tip lift is overlaid for the $\Delta f$ measurements (a-e), showing that there is good agreement between the model and the measurements. The residual of these curves ($\%~\mathrm{Difference}=(\Delta f_{\mathrm{model}}-\Delta f_{\mathrm{data}})/\Delta f_{\mathrm{data}}$) is also shown (k). Finally, in the MIS capacitor model, the $\Delta f$ spectrum is expected to be non-parabolic~\cite{Cowie2022}. The residual of each spectrum in (a) with its parabolic fit shows this slight non-parabolicity for both the data and the model (i). There is some discrepancy between the model and the fit, which is most easily seen given the structured residual (k) and the slightly different trends in the non-parabolicity (i). Overall, however, there is nonetheless good agreement, as the residual percent difference falls within $0-6\%$.

\begin{figure}[!h]
    \centering
    \includegraphics[width=\textwidth]{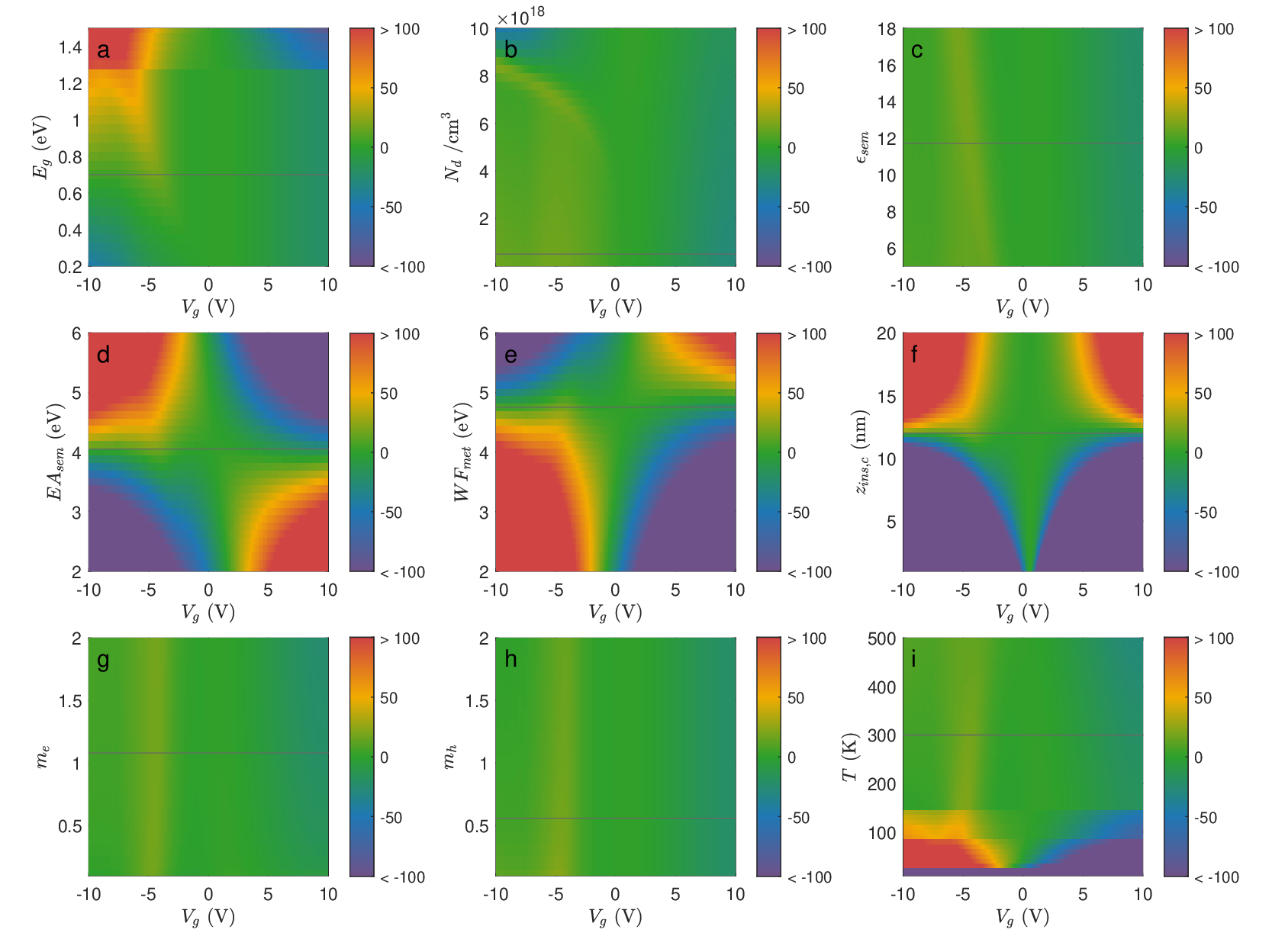}
    \caption{\textbf{MIS model sample parameter sensitivities.} Bias-dependent fm-AFM frequency shift $\Delta f$ residual, constrained to a colour scale between $-100~Hz$ and $100~Hz$, for various sample parameters (band gap $E_g$, donor concentration $N_d$, permittivity $\epsilon_{sem}$, Si electron affinity $EA_{sem}$, tip work function $WF_{met}$, closest tip-sample separation $z_{ins,c}$, electron and hole effective masses $m_e$ and $m_h$, and temperature $T$). The black line shows the parameter used for this work.} 
    \label{fig:Supp_ModelFitSensitivities}
\end{figure}

The sensitivity of the MIS model sample parameters is demonstrated in Figure~\ref{fig:Supp_ModelFitSensitivities}. The bias dependence (horizontal scale) of the $\Delta f$ residual ($\Delta f_{\mathrm{model}}-\Delta f_{\mathrm{data}}$, colour scale) for various parameters (vertical scale) are shown. Note that the colour scale, identical for (a-i), is constrained between $-100~\mathrm{Hz}$ and $100~\mathrm{Hz}$. ($|100|~\mathrm{Hz}$ indicates a very large disagreement between the data and the model and $0~\mathrm{Hz}$ indicates a very good agreement.) Since the colour scales are identical, the various parameter sensitivities can be more easily compared to one another; $EA_{sem}$, for example, is a sensitive parameter whereas $m_e$ and $m_h$ are largely insensitive parameters over `reasonable' values. 

\vspace{-1mm}

\end{document}